\documentstyle{article}
\begin{document}

{
\centerline {\bf {TWO APPROACHES TO ANOMALY-FREE QUANTIZATION}}
\centerline {\bf {OF GENERALLY COVARIANT SYSTEMS}}
\centerline {\bf {ON AN EXAMPLE OF A TWO-DIMENSIONAL STRING}}
\centerline { }

\centerline{ \it S.N.Vergeles.\footnote{e-mail:Vergeles@itp.ac.ru}}
\vspace{3mm}
\centerline { \it {L. D. Landau Institute of Theoretical Physics,
Russian Academy of Sciences.}}
\centerline { \it {142432 Chernogolovka, Russia.}}
\centerline{ }

In this paper we discuss two approaches to anomaly-free quantization of a
 two-dimensional string. The first approach is based on the canonical Dirac
 prescription of quantization of degenerated systems.
At the second approach we "weaken" the Dirac quantization conditions
requiring the solving of first class constraints only in the sense of mean values.
At both approaches there are no states with the indefinite metrics.

\centerline { }
\centerline { }
\centerline { }
\centerline { }
\pagebreak

\centerline {\bf {1. Introduction}}
\centerline { }

The quantum theory of gravity in four-dimensional space-time encounters
fundamental difficulties, which now are not overcome. These difficulties
can be divided on conceptual and computing. The main
conceptual problem is that the Hamiltonian is a linear combination of
a first class constrains. This fact makes unclear a role of time in
gravitaty. The main computing problem consists in nonrenormalizability of
the theory of gravity. The pointed difficulties are closely bound. For
example, an anomaly (central charge) can be present or absent in algebra of first class constraints depending on computing procedure.  In turn the presence or absence
of anomaly in algebra of the first class constraints crucially influences on the resulting physical picture.

The listed fundamental problems are solved successfully for rather simple
models of generally covariant theories in two-dimensional space-time.
Both two-dimensional gravity and string theories belong to such models (see, for example [1-4]
and reference there).

In the present work we develop ideology and technique of anomaly-free
quantization of two-dimensional string. The Hamiltonian of a
two-dimensional bosonic string actually coincides with Hamiltonian of
two-dimensional pure gravity expressed in certain special variables [1,2].
The difference between two-dimensional string and two-dimensional gravity
is that in the theory of two-dimensional string there is an additional
external symmetry relative to abelian Lorentz group in two-dimensional
space-time. However, the ideological basis of anomaly-free quantization
in both models is identical. We explain this idea, using a material of
works [1,2,4].

 Let's consider in two-dimensional
space-time the following system of constraints:
$$
{\cal E}=-{\cal E}_0+{\cal E}_1\approx 0\,,
$$
$$
{\cal E}_0=\frac{1}{2}\,\left(\,(\pi_0)^2+({r^0}^{\prime})^2\right)\,,
 \qquad {\cal E}_1=\frac{1}{2}\,\left(\,(\pi_1)^2+({r^1}^{\prime})^2\right)\,,
   \eqno(1.1a)
$$
$$
{\cal P}={r^{a}}^{\prime}\,\pi_a={r^0}^{\prime}\,\pi_0+
{r^1}^{\prime}\,\pi_1\approx 0  \eqno(1.1b)
$$
Here we use dimensionless values, \ $ r^a(x)$ \
 and \ $ \pi_a(x)), \ a = 0, \, 1 $ \  are canonically conjugated
fields on a one-dimensional manyfold, so the nonzero commutational
relations look like
$$
[r^a(x), \, \pi _ b (y) \,] = i \,\delta ^ a _ b \,\delta (x-y) \eqno(1.2)
$$
The prime or overdot mean the derivatives \ $ \partial/\partial x $ \ or
 \ $ \partial/\partial t $, respectively.

Now we must determine the ground state of the theory. It allows
to perform normal ordering of the operators in the
constraints (1.1). The normal ordering in the constraints can
result to anomalies in the commutators of the constraints. These
anomalies partially destroy the weak equalities (1.1). To determine the
ground state, the fields $r^a$ and $\pi _a$ are expanded in the modes which
arise when solving the Heisenberg equations
$$
i \, {\dot {r}} ^ a = [\, r ^ a, \, {\cal H} \,] \, , \qquad
i \, {\dot {\pi}} _ a = [\, \pi _ a, \, {\cal H} \,] \, ,
$$
$$
{\cal H} = \int \, dx \, {\cal E} \eqno (1.3)
$$
The solution of the equations (1.2), (1.3) can be written in the form
$$
r^a(t, \, x) = \int \,\frac {dk} {2 \,\pi} \, \frac {1} {\sqrt {2 | k |}} \,
\left \{c ^ a _ k \, e ^ {-i \, (| k | \, t-k \, x)} +
c^ {a +} _ k \, e ^ {i \, (| k | \, t-k \, x)} \, \right\} \,,
$$
$$
\pi^a (t, \, x) = -i \,\int \,\frac {dk} {2 \,\pi} \,
 \sqrt {\frac {| k |} {2}} \,
\left \{c ^ a _ k \, e ^ {-i \, (| k | \, t-k \, x)} -
c^{a +} _ k \, e ^ {i \, (| k | \, t-k \, x)} \, \right \} \,,
$$
$$
[\, c ^ a _ k, \, c ^ {b +} _ p \,] = 2 \,\pi \,\eta ^ {ab} \, \delta (k-p) \, \qquad
 [\, c ^ a _ k, \, c ^ b _ p \,] = 0 \eqno(1.4)
 $$
Here \ $ \eta ^ {ab} $ \ (below - \ $ \eta ^ {\mu\nu}\,) = diag (-1,1) $.
We have also the following commutational relations:
$$
 [{\cal H},\, c^a_k\,] = - | k | \, c^a_k \,, \qquad
 [\,{\cal H}, \, c^{a+}_k \,] = |k| \, c^{a+}_k \eqno(1.5)
 $$

Under traditional quantization the operators \ $ c^a _ k $ \ are considered
as the annihilation operators,
and their hermithean conjugated \ $ c ^ {a +} _ k $ \ are considered as
creaton operators.
 The ground state \ $ \vert \, 0 \,\rangle $ \ satisfies the conditions:
$$
c^a_k \,\vert \, 0 \,\rangle = 0 \eqno (1.6)
$$
Normal ordering of the operators \ $ (c^{a+}_k, \, c^a_k \,) $ \ in
values (1.1) means an arrangement of
the creation operators on the left of all annihilation operators.

Let's consider the state
$$
\vert \, k, \, a \,\rangle = c ^ {a +} _ k \, | \, 0 \,\rangle \eqno (1,7)
$$
From commutational relatons (1.5) immediately follows, that
$$
{\cal H} \, \vert \, k, \, a \,\rangle =
( | k | + E _ 0 \,) \,\vert \, k, \, a \,\rangle \,, \eqno (1.8)
$$
where \ $E_0$ \ is the value of the operator \ $ {\cal H} $ \
 for the ground state.
The equality (1.8) implies
 the operator \ $ {\cal H} $ \ is positively defined.

In consequence of (1.4) and (1.6) we have
for scalar product of vectors (1.7):
$$
\langle \, k, \, a \, | \, p, \, b \,\rangle = 2 \,\pi \,\eta ^ {ab} \, \delta (k-p) \eqno (1.9)
$$
Now we see that the metrics in complete space of states is
indefinite.

Let's calculate the commutator \ $ [{\cal E}, \, {\cal P}\,] $,
which can be represented according to (1.1) as a sum of two terms:
$$
[\,{\cal E}(x),\,{\cal P}(y)\,]=
-[\,{\cal E}_0(x),\,{r^0}^{\prime}\,\pi_0(y)\,]+
[\,{\cal E}_1(x),\,{r^1}^{\prime}\,\pi_1(x)\,]   \eqno(1.10)
$$
According to (1.2) the both commutators in the right hand side of Eqs.(1.10)
 coincide up to replacement of index \ $ "a" $. These commutators
are proportional (up to the ordering) to quantities \ $ {\cal E} _ 0 $ \
and \ $ {\cal E} _ 1 $ \ , respectively. The normal ordering of the
operators in considered
commutators, as is known, results in anomalies.

Indeed, from the commutational relations (1.4) it follows that
the correspondenses  \ $ c ^ 0 _ k\leftrightarrow c ^ {1 +} _ k \,, \
 c^{0 +} _ k\leftrightarrow c ^ 1 _ k $ \
 give the isomorphism of Heisenberg algebras  \ $ H _ 0 $ \ and \ $ H _ 1 $ \
 with generators \ $ (c ^ 0 _ k, \, c ^ {0 +} _ k \,) $ \ and
 \ $ (c ^ {1 +} _ k, \, c ^ 1 _ k \,) $ respectively. In this case
the normal ordering of the operators in algebra \ $ H _ 1 $ \ is
mapped by
specified isomorphism to antinormal ordering in algebra \ $ H _ 0 $ \ .

 As is known, in such case the normal and antinormal
 orderings result in anomalies differing only in sign.
Hence, the contribution of the first commutator in
anomaly in the right hand side of Eq.(1.10) will be \ $ (-A) $ \ and
of the second commutator will be \ $ (A) $. But
as in front of the first commutator in
(1.10) there is a sign " minus ", the anomaly in (1.10) is equal
 \ $ - (-A) + A = 2 \, A $.

Now we pass to another point of view.

In the work [1] Jackiw states that the condition (1.8)  of positivity of operator
 \ $ {\cal H} $ \ is not
 necessary in the considered theory. The initial requirement of
the theory is that the weak equalities (1.1) are valid. Therefore we can
 refuse the conditions of quantization (1.6) and replace them
by the following:
$$
c^{0 +} _ k \,\vert \, 0 \,\rangle = 0 \,, \qquad
 c^ 1 _ k \,\vert \, 0 \,\rangle = 0 \eqno (1.11)
$$

Under the quantization conditions (1.11) the bases of complete Fock  space
of the theory consists of the following vectors:
$$
c^ 0 _ {k _ 1} \, \ldots \, c ^ 0 _ {k _ m} \, c ^ {1 +} _ {p _ 1} \,
\ldots \, c ^ {1 +} _ {p _ n} \, \vert \, 0 \,\rangle  \eqno(1.12)
$$

From the commutational relations (1.4) it follows that the
scalar product of states (1.12) is positively defined. Moreover, in
the algebra of the operators
(1.1) {\it {there is no anomaly.}}

Indeed, under conditions (1.11) the normal ordering consists in
arrangement of the operators \ $ (c ^ 0 _ k, \, c ^ {1 +} _ k \,) $ \
on  the left of all operators
 \ $ (c ^ {0 +} _ k, \, c ^ 1 _ k \,) $. It means the normal ordering in both
Heisenberg algebras \ $ H _ 0 $ \ and \ $ H _ 1 $ \ . Now at the normal
ordering the both commutators in (1.10) give in
anomaly the same contribution, which is equal  to \ $ A $. Since
in front of the first commutator in
the right hand side of Eq.(1.10) there is a sign "minus"
 the complete anomaly
 appears equal to
 \ $ -A + A = 0 $.

The absence of anomaly in algebra of the operators
\ $({\cal E}, \, {\cal P}\,) $ \
 enables to satisfy
all weak equalities \ ${\cal E} \approx 0 \,, \ {\cal P} \approx 0 $.
In [1] two physical states annulating the all
operators \ $ {\cal E} $ \ and \ $ {\cal P} $ \ are given:
$$
\Psi _ {gravity} (r ^ a) = \exp\pm\frac {i} {2} \, \int \, dx \,\varepsilon _ {ab} \,
r^a \, {r ^ b} ^ {\prime}
$$

In [2,4] the continuous families of states parametrized by one real parameter which solve the all
constraints (1.1) are also considered.

In this work we reconsider the well known quantization conditions
for relativistic bosonic string in two-dimensional space -
time. As a result we formulate two another approaches to quantizarion of considered model. The both approaches lead to the Virasoro algebra in which the central charge is absent. We use complete space of states with the positively defined
scalar product. The complete space of states is decomposed to tensor
product of physical states space and gauge states spaces. Such
decomposition make possible a calculation of matrix elements for dynamic
variables separately relative to physical states and states describing gauge degrees of
freedom. This fact allows to change the Dirac quantization prescription for systems containing the first class constraints [5]. The arising
 quantization rule can be useful for quatization of more complicated
degenerated systems for example such as four-dimensional gravity.

\centerline { }
\centerline { }
\centerline {\bf {2. Relativistic bosonic string}}
\centerline {\bf {in two-dimensional space-time}}
\centerline { }

Let \ $ X ^ {\mu} \, \ \mu = 0, \, 1 $ \ be the coordinates
in two-dimensional Minkovski space. Consider
the Nambu action for bosonic string:
$$
S = -\frac {1} {l ^ 2} \, \int \,\sqrt{-g} \, d ^ 2\xi = \int \, d\tau \, {\cal L} \eqno(2.1)
$$
Here \ $ \xi ^ a = (\tau, \, \phi) $ \ are the parameters of the world
sheet of the string and
$$
g = Det \, g _ {ab} \, \qquad g _ {ab} = \eta _ {\mu\nu} \,
\frac {\partial X^{\mu}} {\partial\xi ^ a} \,
\frac {\partial X ^ {\nu}} {\partial\xi ^ b}
$$
The parameter \ $ \tau $ \ is time like and \ $ \phi $ \ is spacial.
Next the partial derivatives \ $ \partial/\partial\tau $ \
and \ $ \partial/\partial\phi $ \ will be denoted by an over dot
and a prime, respectively. It is easy to show, that
the generalized momenta \ $ \pi _ {\mu} =
\partial {\cal L} /\partial {\dot {X}} ^ {\mu} $ \ satisfy
the conditions:
$$
{\cal E} = \frac {l ^ 2} {2} \, \pi _ {\mu} \, \pi ^ {\mu} + \frac {1} {2 \, l ^ 2} \,
{X ^ {\mu}} ^ {\prime} \,
X ' _ {\mu} \approx 0 \,,
$$
$$
{\cal P} = {X ^ {\mu}} ^ {\prime} \, \pi _ {\mu} \approx 0 \eqno(2.2)
$$
The quantities \ $ {\cal E} (\phi) $ \ and \ $ {\cal P} (\phi) $ \ exhaust
the all first class constraints. The hamiltonian of the system
$$
{ \cal H} = \int \, d\phi \,\pi _ {\mu} \, {\dot {\phi}} ^ {\mu} - {\cal L} \approx 0
$$
is also equal to zero.

 Therefore, following to Dirac, we must use the generalized
hamiltonian which is an arbitrary linear combination of the first class constraints (2.2):
$$
{\cal H} _ T = \int \, d\phi \, (v \, {\cal P} + w \, {\cal E}) \eqno(2.3)
$$
The equations of motion can be obtained with the help of the variational principle
$$
\delta \, S = \delta \, \{\,\int \, d\tau \, (\,\int \, d\phi \,\pi _ {\mu} \,
{ \dot {X}} ^ {\mu} - {\cal H} _ T) \, \}=0           \eqno(2.4)
$$
In the case of an open string, when the parameter \ $ \pi $ \ changes from zero to
 \ $ \pi $ , the variational principle (2.4)
gives besides the equations of motion the boundary conditions
$$
( v \,\pi _ {\mu} + w \,\frac {1} {l ^ 2} \, X ' _ {\mu}\,) \, \vert _ {\phi =
 0, \, \pi} = 0      \eqno(2.5)
$$
Usually the boundary conditions (2.5) are replaced by conditions
$$
v\vert _ {\phi = 0, \, \pi} = 0 \,, \qquad
X ' _ {\mu} \vert _ {\phi = 0, \, \pi} = 0     \eqno(2.6)
$$
For the closed string instead of the boundary condition there is the periodicity
condition.

Further we consider an open string.

The first step of quantization consists in definition of commutational relations
for generalized coordinates and momenta:
$$
[ \, X ^ {\mu} (\phi), \, \pi ^ {\nu} (\phi ') \,] = i \,\eta ^ {\mu\nu} \, \delta (\phi-\phi ')
   \eqno(2.7)
$$
The commutational
relations (2.7) and the boundary conditions (2.6) are satisfied if
$$
X ^ {\mu} (\phi) = \frac {l} {\sqrt {\pi}} \, \left (x ^ {\mu} + i \,
\sum _ {n\neq 0} \, \frac {1} {n} \, \alpha ^ {\mu} _ n \,\cos {n\phi} \right) \, ,
$$
$$
\pi ^ {\mu} (\phi) = \frac {1} {\sqrt {\pi} \, l} \, \sum _ {n} \,
\alpha ^ {\mu} _ n \,\cos {n\phi} \,       \eqno(2.8)
$$
and the elements \ $ (x ^ {\mu}, \, \alpha ^ {\mu} _ n) $ \ satisfy
 the following commutational relations:
$$
[ \, x ^ {\mu}, \, \alpha ^ {\nu} _ n \,] = i \,\eta ^ {\mu\nu} \, \delta _ n \,, \qquad
[ \, x ^ {\mu}, \, x ^ {\nu}\,] = 0 \,,
$$
$$
[ \, \alpha ^ {\mu} _ m, \, \alpha ^ {\nu} _ n \,] =
 m \,\eta ^ {\mu\nu} \, \delta _ {m + n}       \eqno(2.9)
$$
Since the quantities (2.8) are real then
$$
x ^ {\mu +} = x ^ {\mu} \, , \qquad      \alpha ^ {\mu +} _ n =
\alpha ^ {\mu} _ {-n}      \eqno(2.10)
$$
The constraints (2.2) can be represented as follows:
$$
( {\cal E} \pm {\cal P}) (\phi) = \frac {1} {2} \, (\xi ^ {\mu} _ {\pm} (\phi)) ^ 2 \,,
   \eqno(2.11)
$$
where
$$
\xi ^ {\mu} _ {\pm} (\phi) = \frac {1} {\sqrt {\pi}} \, \sum _ n \,\alpha ^ {\mu} _ n \,
\exp\mp i \, n \,\phi      \eqno(2.12)
$$
From here it is seen, that \ $ {\cal E} - {\cal P} $ \ is obtained
from \ $ {\cal E} + {\cal P} $ \ by replacement \ $ \phi \longrightarrow -\phi $.
 This fact simplifies
the further analysis, as the quantity \ $ {\cal E} + {\cal P} $ \ in
interval \ $ -\pi\leq\phi\leq\pi $ \ contains the information about
quantities \ $ {\cal E} \pm {\cal P} $ \ in  interval
 \ $ 0\leq\phi\leq\pi $. Therefore the Fourier components
 $$
 L_n = \frac {1} {2} \, \int ^ {\pi} _ {-\pi} \, d\phi \, ({\cal E} +
 {\cal P}) \, \exp \, i \, n \,\phi     \eqno(2.13)
 $$
are equivalent to the set of quantities (2.2) at \ $ 0\leq\phi\leq\pi $. According to
(2.11) - (2.13) we have
$$
L_n = \frac {1} {2} \,: \,\sum _ m \,\alpha ^ {\mu} _ {n-m} \, \alpha _ {\mu \, m} \,:
      \eqno(2.14)
$$
The sense of the ordering operation in (2.14) is defined by a method of quantization.

Let's write out also the expressions for momentum and angular momentum
 of the string:
$$
P ^ {\mu} = \int ^ {\pi} _ 0 \, d\phi \,\pi ^ {\mu} \, , \qquad
 J ^ {\mu\nu} = \int ^ {\pi} _ 0 \, d\phi \, (X ^ {\mu} \, \pi ^ {\nu} -
 X ^ {\nu} \, \pi ^ {\mu})       \eqno(2.15)
 $$
With the help of (2.6) and (2.7) we can directly verify, that
$$
[ \, P ^ {\mu}, \, {\cal H} _ T \,] = 0 \,, \qquad [\, J ^ {\mu\nu}, \, {\cal H} _ T \,] = 0
$$
It means, that the momentum and angular momentum of the string
are conserved.

Under standard  quantization the ground state
 \ $ \vert \, 0 \,\rangle $ \ satisfies the conditions
 $$
 \alpha ^ {\mu} _ m \,\vert \, 0 \,\rangle = 0 \,, \qquad m\geq 0 \eqno (2.16)
$$
The complete space of states has the orthogonal basis:
 $$
\alpha ^ {\mu _ 1} _ {m _ 1} \, \ldots \,\alpha ^ {\mu _ s} _ {m _ s} \, \vert \, 0 \,\rangle \,,
 \qquad m _ i < 0        \eqno(2.17)
$$

Thus, the all \ $ \alpha ^ {\mu} _ m $ \ are linear
  operators in complete space of states.
 From (2.9) and (2.16) it follows, that the metrics
in the space of states (2.17)
is indefinite. The ordering in (2.14) means, that the operators
 \ $ \alpha ^ {\mu} _ m $ \ with \ $ m < 0 $ \  are placed
  on the left of the all operators \ $ \alpha ^ {\mu} _ n $ \ with \ $ n\geq 0 $.
 The such ordering results in
the Virasoro algebra contains the anomalies:
$$
[ \, L _ n, \, L _ m \,] = (n-m) \, L _ {n + m} + \frac {1} {12} \, D (n ^ 3-n)
       \eqno(2.18)
$$
Here \ $ D $ \ is the dimension of \ $ x $ - space. In our
case \ $ D =2$ . Therefore
the annihilation of the operators
 \ $ L _ n $ \ with \ $ n\geq 0 $ is maximum, that can be reached.
As a result
the theory is self-consistent only for \ $ D = 26 $. The detailed
study of problems,
arising under the quantization (2.16), can be found in [4].

Now we shall state the way of quantization of two-dimensional string,
 which
results in the self-consistent string theory [1-4]. This quantization method of
 the string
is similar to the quantization method applied by Dirac to
 electromagnetic
field (see. [7], and also Appendix).

Let's introduce the designations
$$
x_{\pm} = x ^ 0\pm x ^ 1 \,, \qquad \alpha ^ {(\pm)} _ m =
\alpha ^ 0 _ m\pm\alpha ^ 1 _ m
  \eqno (2.19)
$$
From (2.9) we obtain:
$$
[ \alpha ^ {(+)} _ m, \, \alpha ^ {(+)} _ n \,] =
[ \alpha ^ {(-)} _ m, \, \alpha ^ {(-)} _ n \,] = 0 \,, \qquad
[ \alpha ^ {(+)} _ m, \, \alpha ^ {(-)} _ n \,] =-2m \,\delta _ {m + n}
$$
$$
[ x _ +, \, x _ - \,] = 0 \,, \qquad \ [x _ +, \, \alpha ^ {(+)} _ n \,] =
[ x _ -, \, \alpha ^ {(-)} _ n \,] = 0 \,,
$$
$$
[ x _ +, \, \alpha ^ {(-)} _ n \,] =-2i \,\delta _ n \,,      \qquad \
 [x _ -, \, \alpha ^ {(+)} _ n \,] =-2i \,\delta _ n        \eqno (2.20)
 $$
Let's write out the Virasoro operators in variables \ $ \alpha ^ {(\pm)} $:
$$
L _ n = -\frac {1} {2} \, \,\sum _ m \,\alpha ^ {(+)} _ {n-m} \, \alpha ^ {(-)} _ m \,
  \eqno (2.21)
  $$

By definition the ordering in (2.21) means either the elements
 \ $ \alpha ^ {(+)} $ \ are arranged on the left of the all elements \ $ \alpha ^ {(-)} $
 or the elements \ $ \alpha ^ {(-)} $ \ are arranged on
 the left of the all
elements \ $ \alpha ^ {(+)} $. Both these orderings are
equivalent. Indeed
$$
\sum _ m \,\alpha ^ {(-)} _ (-m) \,\alpha ^ {(+)} _ m =
\sum _ m \,\alpha ^ {(+)} _ m \,\alpha ^ {(-)} _ (-m)
+ 2 \,\sum _ m \, m \,,
$$
It is possible to consider the last sum as equal to zero, since it is represented in the form
 \ $ \zeta (-1) -\zeta (-1) $, where \ $ \zeta (s) $ \ is
the Rieman  zeta-function. As is known the zeta-function
 $$
 \zeta (s) \equiv\sum ^ {\infty} _ {n = 1} \, n ^ {-s}  \eqno(2.22)
 $$
has the unique analytical continuation to the point \ $ s = -1 $ \ and
 \ $ \zeta (-1) = -1/12 $.

The following problem consists in the definition of vector space of states, in
which the dynamic variables of the system act as linear operators.

In the beginning we apply the Dirac prescription for quantization of
our model,
marking the chosen way by the index \ $ D $. We represent the complete space of
states \ $ H_{CD} $ \ as the tensor product of the gauge space
of states \ $H_G$ \
and physical space of states \ $H_{PD} $:
$$
H _ {CD} = H_G\otimes H_{PD} \eqno(2.23)
$$

The space \ $ H_G $ \  is generated by its vacuum vector \ $ | \, 0; \, G
\,\rangle $, which is determined by the following properties:
$$
\alpha^0_{-m} \, | \, 0; \, G \,\rangle = 0 \,,
\ \ \ \alpha ^ 1 _ m \, |\, 0; \, G \,\rangle = 0 \,, \ m> 0 \,,
\ \ \ \langle \, 0; \, G \, | \, 0;
\, G \,\rangle = 1 \eqno (2.24)
$$
The basis of the space \ $ H _ G $ \ consists of vectors of the form
$$
\alpha ^ 0_{m _ 1} \ldots\alpha ^ 0 _ {m
_ s} \, \alpha ^ 1 _ {-n _ 1} \ldots \alpha ^ 1 _ {-n _ r} \, | \, 0; \, G
\,\rangle \,, \ \ \ m _ i > 0 \,, \ n _ i > 0 \eqno (2.25)
$$
Thus, \ $H_G$ \ is the Fock space with positively
defined scalar product.

The basis in the Dirac physical space of states \ $H_{PD}$ \ consists
 of
two series \ $ | \, k\pm \,\rangle_D$ \ with the
following properties (\ $
\alpha ^ {(\pm)}_0\equiv p_{\pm} $ \ ):
$$
p_+ \, | \, k- \,\rangle_D = 2k
\, | \, k- \,\rangle_D \,, \ \ \ p _-\, | \, k + \, \rangle_D = 2k \, | \,
k + \, \rangle_D \,, \eqno(2.26)
$$
$$
\alpha^{(-)}_m \, | \, k- \,\rangle_D = 0 \,, \ \ \ \alpha^{(+)}_m \, | \,
k + \, \rangle_D = 0 \,, \ \ m = 0, \, \pm 1, \ldots    \eqno(2.27)
$$
and
$$
\langle \, k\mp \, | \, k^{\prime} \mp \,\rangle_D =
k \,\delta (k-k^{\prime} \,) \,, \ \ \ \langle \, k- \, | \, k ^ {\prime} +
\, \rangle_D = 0   \eqno(2.28)
$$
Here \ $ k $ \ is continuous real parameter, \ $ 0 < k < + \infty $. The
quantization conditions (2.27) were applied in works [2,4].
Earlier the similar
quantization conditions were applied by Dirac in electrodynamics [7].

From the given definitions it is seen, that in complete space of states
the scalar product is positively defined.

 Let's emphasize, that variables \ $ \alpha ^ {\mu} _ n \,, \ n\neq 0 $,
being the linear operators in space \ $ H _ G $, are not operators in
the space \ $ H_{PD} $. Indeed, the action of the variables \ $
\alpha^{(-)} _ n $ \ on the vectors \ $ | \, k + \, \rangle_D $, and also
action of the variables \ $ \alpha^{(+)} _ n $ \ on the vectors \ $ | \, k-
\,\rangle_D $ \ is not defined, if \ $ n\neq 0 $. Nevertheless from
the given definitions it follows, that the action of some combinations
(generally speaking, nonlinear) of the variables \ $ \alpha^{\mu}_n $ \
with \ $ n\neq 0 $ in the space \ $ H_{PD} $ \ is determined correctly. For
example, owing to (2.20) and (2.27), we have:
$$
 \alpha ^ {(+)} _ m
\,\alpha ^ {(-)} _ {-m} \, | \, k + \, \rangle _ D = -2m \, | \, k +
\rangle_D \,, \ \ \ \alpha ^ {(-)} _ m \,\alpha ^ {(+)} _ {-m} \, | \, k-
\,\rangle_D = -2m \, | \, k- \,\rangle _ D \eqno (2.29)
$$
Since
$$ L _ n =
-\frac {1} {2} \, \sum _ m \,\alpha ^ {(+)} _ {n-m} \, \alpha ^ {(-)} _ m =
-\frac {1} {2} \, \sum _ m \,\alpha ^ {(-)} _ {n-m} \, \alpha ^ {(+)} _ m
\,,
$$
and as the consequence of (2.27) the equalities
$$
L _ n \, | \, p
\,\rangle_D = 0 \,, \qquad | \, p \,\rangle _ D\in H _ {PD} \, \eqno (2.30)
$$
take place. Thus, all physical states are annihilated by the all
Virasoro operators. The equalities (2.30) mean, that under quantization
(2.24) - (2.28) the Virasoro algebra has no anomaly.
$$
[ L _ n, \, L _ m \,] = (n-m) \, L _ {n + m} \eqno (2.31)
$$
 Here we have a situation, which is common for all degenerated
systems. In degenerated systems there is a set of first class constraints \
$ \chi_m $. If evident resolution of these constraints is inexpedient,
then according to Dirac prescription the
following conditions
$$ \chi_m \, | \, p \,\rangle_D = 0
\eqno(2.32)
$$
are imposed on the physical states.
It is obvious, that owing to (2.32) the physical states can
not depend on gauge degrees of freedom, the changes of which are generated
by the constraints \ $ \chi _ m $. Thus the variables, describing the gauge
degrees of freedom can not be , generally speaking, the linear operators in
the space of physical states. On the other hand, according to Dirac
prescription it is necessary to consider only the
physical states.

Let's consider on an example of investigated model the paradox which arise
when the Dirac procedure is realized. We have a set of the oscillator
variables \ $ \alpha^{\mu}_m \,, \ m\neq 0 $, and also the coordinate and
momentum variables \ $(x^{\mu}, \, p^{\mu}\,)$. Under the Schrodinger
 quantization the complete space of states \ $ H_C $ \ is decomposed to the
tensor product of the spaces \ $ H_G $ \ and \ $ H_P $: \ $ H_C =
H_G\otimes H_P $. Here the space \ $ H_G $ \ is as in (2.24) -
 (2.25).
%4 €...‹Ÿ 1999 ƒ.

The space \ $ H _ P $ \ in the Schrodinger representation is, for
 example, the space of functions of two variables \ $ (x ^ 0, \, x ^ 1 \,)
$, on which there is a positively determined Lorentz - invariant scalar
product. By this definition in the space \ $ H _ P $ \ there are no states
which are annihilated by the operators \ $ \alpha^{(\pm)} _ m \,, \ m\neq 0
$. Moreover, also in complete space \ $ H_C $ \ there are no states which
are annihilated by the operators \ $ \alpha ^ {(\pm)}_m \,, \ m\neq 0 $.
One can add to the space \ $ H _ G $ \ the states \ $ | \, \pm \,\rangle $,
on which the operators \ $ \alpha^{(\pm)}_m $ are equal to zero.  In the
extended space we shall define the scalar product as follows:
$$ \langle
\,\pm \, | \,\pm \,\rangle = 1 \,, \qquad \langle \, + \, | \,- \,\rangle =
0 \,,
$$
$$
\langle \,\pm \, | \, G \,\rangle = 0 \,, \qquad | \, G
\,\rangle\in H _ G \eqno (2.33)
$$
Further we define
$$ | \, k\pm \,\rangle
_ D = | \, \pm \,\rangle \,\otimes \, | \, k\pm \,\rangle \,,
$$
where
$$
| \, k\pm \,\rangle \,\in H _ P \,, \qquad p _ {\pm} \, | \, k\mp \,\rangle =
2k \, | \, k\mp \,\rangle \eqno (2.34)
$$
Then the set of vectors \ $ \{\,
| \, k\pm \,\rangle _ D \} $ \ form the basis of the space \ $H_{PD}$.

Let's pay attention to the reason by which the parameter \ $ k $ \ in
(2.27), (2.28) and (2.34) is positive. This follows from the requirement of a
positivity of scalar product in the space \ $ H_P $. Indeed, the space \ $
H _ P $ \ is the space of states of a massless bose-particle. As is well
known (see [8], Chapter 3, \S 2) in such space the positively defined and
Lorentz - invariant scalar product exists only for positive (or negative)
energies.

From above consideration it is seen, that any vector from the space \ $ H
_ {PD} $ \ does not belong to initial normalizable space \ $ H _ C $. But
just in the space \ $ H_C $ \ the initial variables \ $ \alpha^{\mu}_m \,,
\ m\neq 0 $ \ are the well defined bose creation and annihilation
operators.

 Thus, the Dirac prescription imply the variables \ $ \alpha^{\mu} _ m $ \
with \ $ m\neq 0 $ \ are not the operators in the physical space \ $ H_{PD}
$. The variables \ $ \{\alpha^{\mu} _ m \,, \ x^{\mu} \, \} $ \ are
generators of the associative noncommutative involutive algebra \ $ {\cal
A} $ \ with unit over the complex numbers (see [4]). The generators \ $
\{\alpha^{\mu}_m, \, x^{\mu} \, \} $ \ satisfy only the relations
(2.9). The generators \ $ \{x^{\mu}, \, p^{\mu} \, \} $ \ are the operators
in the space of physical states \ $ H_{PD} $. In the space \ $ H_{PD} $ \
there is the basis \ $ | \, k\pm \,\rangle _ D $, for which the relations
(2.27) take place.

The variables $ \alpha^{\mu}_m $ \ with \ $ m\neq 0 $ \ are not the
operators in the space of physical states. Nevertheless, the observable
quantities can depend on these variables so, that the matrix elements of
observable quantities relative to vectors from the space of physical states
are determined correctly. Just such situation has a place in quantum
electrodynamics for Dirac quantization [7] (see also Appendix).

Let's continue the study of our model.

From the definitions (2.15) we obtain the following formulae:
$$
( \exp \, i\omega \, J^{01}) \, \alpha^{(\pm)}_m \,(\exp-i\omega \, J^{01})=
(\exp\pm\omega) \, \alpha^{(\pm)}_m \,,
$$
$$
( \exp \, i\omega \, J^{01}) \, x_{\pm}\, (\exp-i\omega \, J^{01}) =
(\exp\pm\omega) \, x_{\pm} \eqno (2.35)
$$
and
$$
( \exp \, ia _ {\mu} P^{\mu}) \, x_{\pm}\, (\exp-ia_{\mu} P^{\mu}) =
x _ {\pm} + \frac {\sqrt {\pi}} {l} \, a_{\pm} \,
$$
$$
( \exp \, ia _ {\mu} P^{\mu}) \, \alpha^{(\pm)}_m \, (\exp-ia_{\mu} P^{\mu}) =
\alpha ^ {(\pm)} _ m \eqno (2.36)
$$
Here $\omega $ and $a^\mu $ are arbitrary real numbers. It is evident from
Eqs. (2.29) and (2.30) that translations and Lorentz transformations
conserve the condition (2.23).

According to Eq. (2.35)
$$
p_{\pm} \, (\exp-i\omega \, J ^ {01}) = (\exp \,\omega)\,
 (\exp-i\omega \, J ^ {01}) \, p_{\pm}
  \eqno (2.37)
$$
Let us formally act with the equalities (2.37) on the states $\left|
k\mp\right\rangle $, accordingly. As a result of Eq. (2.26) we obtain
$$
p_{\pm} \, (\exp-i\omega \, J^{01}) \, \vert \, k \mp\,\rangle = 2k \,
e^{\pm\omega} \, ( \exp-i\omega \, J ^ {01}) \, \vert \, k\mp \,\rangle
\eqno (2.39)
$$
The last equality makes it possible to determine the action
of the operators $(\exp -i\omega J^{01})$ on the physical states as
follows:
$$
( \exp-i\omega \, J ^ {01}) \, \vert \, k \mp\,\rangle =
f^{\mp}_{\omega} \, \vert \, (\exp{-\omega}\,J^{01})\,|\,k\mp \,\rangle
$$
Here $f^{\mp}_\omega $ are some complex numbers different from zero. Since
the scalar product on physical state vectors is defined in a
Lorentz-invariant manner according to (2.28) so $\left| f^{\mp}_\omega
\right| =1$. From Eq. (2.39) it is evident that one can assume
$$
k> 0
\eqno (2.40)
$$

According to Eqs. (2.8) and (2.15)
$$
P ^ {\mu} = \frac {\sqrt {\pi}} {l} \, \alpha ^ {\mu} _ 0 =
\frac {\sqrt {\pi}} {2 \, l} \,
\{ (\delta^{\mu}_0 + \delta^{\mu} _ 1) \, p _ ++
 (\delta^{\mu}_0-\delta^{\mu} _ 1) \, p _ - \}
$$
Therefore from (2.26) and (2.27) we obtain:
$$
P^{\mu} \, \vert\, k\mp\,\rangle_D =
\frac{\sqrt {\pi}} {l}\, k_{\pm}^{\mu} \, \vert \, k\mp \,\rangle_D \,,
 \qquad k_{\pm}^{\mu} = (k, \, \pm k) \eqno (2.41)
$$

Thus, as a result of the above-described quantization procedure of
two-dimensional string there arises a system similar to a massless
quantum particle in two-dimensional space-time.

\centerline { }
\centerline { }
\centerline {\bf {3. The other way of anomaly-free quantization}}
\centerline {\bf {of a two-dimensional string.}}
\centerline { }

In the previous section we have emphasized the difficulties which arise
under Dirac quantization of degenerated systems. These difficulties
in general are
reduced to the following problems:

1) The problem of normalizability of physical state vectors.

2) The impossibility of interpretation of some
initial dynamic variables as linear operators in the space of physical
states.

These problems are successfully solved in relatively simple models such as electrodynamics and two-dimensional string. However, the complete solution of these problems in more
complicated models such as four dimensional gravity is extremely difficult.

 We propose in this section the alternative way of
quantization of a two-dimensional string. The idea of this quantization
consists in some weakening of the Dirac conditions (2.32) by replacement
them with conditions:
$$
\langle \, P \, | \,\chi_m \, | \, P \,\rangle_G =
0 \eqno (3.1)
$$
Here the index \ $ P $ \ number any physical state. The index \ $ G $ \
indicates that an averaging in (3.1) goes only in gauge degrees of freedom.
The quantization conditions (3.1) are similar in some sense a) to the Gupta-Bleuler conditions in electrodynamics, when the equality \
$ \partial_{\mu} A^{\mu} = 0 $ \ takes place only in the sense of mean
value; b) to the quantization conditions in the usual string theory, when
the generators of Virasoro algebra satisfy conditions \ $ L_n = 0 $ \
also only in the sense of mean value.

It is important that calculation of the mean value in (3.1) is performed
relative to physical states.

The difference of offered here quantization from Gupta-Bleuler
and standard string quantizations is that at our approach the scalar
product in
complete space of states is positively defined. It is shown below,
that this fact allows to carry out the
anomaly-free quantization of a two-dimensional string.

Insted of the Dirac
selfconsistency conditions \ $[\chi_m, \, \chi_n \,]=c_{mn}^l\,\chi_l$ ,
 now we have:
$$
\langle\,P\,|\,[\chi_m,\,\chi_n \,] \,|\, P \,\rangle_G=0 \eqno (3.2)
$$
Explain the physical sense of the conditions (3.2). As is known
Hamiltonian of a generally covariant system has the form \ $ {\cal H}_T = \sum
\, v _ m\chi_m $. Assume, that at the moment of time \ $ t $ \ the
conditions (3.1) take place. In the infinitely close moment of time \ $ t +
\delta \, t $ \ the constraint \ $ \chi _ n $ is equal to
$$
\chi _ n (t +
\delta t) = \chi _ n (t) + i \,\delta t \,\sum _ m \, v _ m \, [\chi_m, \,
\chi_n \,] (t)
$$
Thus the selfconsistency conditions (3.2) provide the
equalities (3.1) at any moment of time.

We believe, that the complete space of states, in which the initial
dynamic variables (2.9) - (2.10) act, is the space \ $ H_C $, determined in
the previous section:
$$
H _ C = H _ G\otimes H _ P \eqno (3.3)
$$
Here
the spaces \ $ H _ G $ \ and \ $ H _ P $ \ are defined according to (2.24),
(2.25) and (2.34). Let in the space \ $ H _ P $ \ the basis \ $ | \, k
\,\rangle = | \, k ^ 0, \, k ^ 1 \,\rangle $ \ be such, that
$$
p ^ {\mu}\, | \, k \,\rangle = k ^ {\mu} \, | \, k \,\rangle \eqno (3.4)
$$
For the further calculations it is necessary to define a suitable ordering of
the operators. Next we shall use the ordering
$$
L _ 0 = \frac{1}{2} \, p^{\mu} \, p_{\mu} - \sum_{m> 0} \,
(\alpha^0 _ m \,\alpha^0 _ {-m} -
\alpha^1_{-m} \, \alpha^1_m \,) \,, \eqno (3.5)
$$
which is equivalent to the ordering (2.21).

We think that in the examined model the most convenient physical states
satisfying conditions (3.1), are the states, which are coherent relative
to gauge degrees of freedom. Let's consider in the space \ $ H_G $ \ the
coherent state
$$
| \, z; \, G \,\rangle =
\prod_{m> 0} \, \exp\left \{-\frac {| \, z^0_{-m} \,
|^2 + | \, z^1 _ m \, |^2} {2m}
+ \frac{1}{m} \, (z^0 _ {-m} \, \alpha^0_m+z^1_m \,\alpha^1_{-m}\,
\, \right \} \, | \, 0; \, G \,\rangle \eqno (3.6)
$$
Here \ $ z^{\mu} _ m \,, \ m\neq 0 $ \ are complex numbers. Below we
assume by definition that \ $ z^{\mu} _ 0 = k^{\mu} $ \ and \ $ {\bar
{z}}^{\mu} _ m = z^{\mu}_{-m} $. The bar above means the complex
conjugation. As a consequence of (2.9) and (2.24) we have:
$$
\langle\,z;\,G\,|\,z;\,G\,\rangle=1\,,
$$
$$
\alpha^0_{-m}\,|\,z;\,G\,\rangle=z^0_{-m}\,|\,z;\,G\,\rangle\,,
$$
$$
\alpha^1_{m}\,|\,z;\,G\,\rangle=z^1_{m}\,|\,z;\,G\,\rangle\,,
 \ \ \ m>0          \eqno(3.7)
$$

Let's introduce a designation \ $ | \, z \,\rangle $ \ for the state
$$
 |\,z \,\rangle = | \, z; \, G \,\rangle\otimes | \, k \,\rangle\in H _ C
 \eqno (3.8)
$$
From (3.4), (3.7) and (3.8) it follows, that
$$
 \alpha^0 _ {-m} \, | \, z \,\rangle = z ^ 0 _ {-m} \, | \, z \,\rangle
 \,, \ \ \alpha ^ 1 _ {m} \, | \, z \,\rangle = z ^ 1 _ {m} \, | \, z
\,\rangle \,, \ \ m\geq 0 \eqno (3.9)
$$
We call the set of complex numbers
\ $ \{\, z ^ {\mu}_m \, \} $ \ as parameters of the state \ $ | \, z
\,\rangle $. A state is called physical and designated by \ $ | \, z
\,\rangle_P $, if its parameters satisfy the equations:
$$
L_n(z)\equiv\frac{1}{2}\,\sum_m\,z^{\mu}_{n-m}\,z_{\mu m}=0
  \eqno(3.10)
$$

We put \ $ z^{(\pm)}_n = z ^ 0 _ n\pm z^1_n $ and
$$
z^{(\pm)}(\phi)=\frac{1}{\sqrt{2\pi}}\,\sum_n\,e^{-in\phi}
\,z^{(\pm)}_n\,,
$$
$$
L(\phi)=\frac{1}{2\pi}\,\sum_n\,e^{-in\phi}\,L_n=
-\frac{1}{2}\,z^{(+)}(\phi)\,z^{(-)}(\phi)
$$
The functions \ $z^{(\pm)}(\phi)$ \ are real. Equations
(3.10) are equivalent to the
following one:
$$
L(\phi)=-\frac{1}{2}\,z^{(+)}(\phi)\,z^{(-)}(\phi)=0 \eqno(3.10')
$$

Impose also the gauge invariant conditions
$$ -z^{(+)} _ 0 \, z ^ {(-)} _ 0\equiv k ^ {\mu} \, k _ {\mu} = 0 \eqno
(3.11)
$$
Further it is supposed, that all solutions of the equations
(3.10) satisfy condition (3.11). Therefore the basis in the space \ $
H_P $ \ can be given according to (2.34), and the scalar product
According to (2.28).

 From the formulas (3.5), (3.9), (3.10) it follows
immediately, that in considered case the conditions (3.1) take place:
$$
\langle \, z \, | \, L _ n \, | \, z \,\rangle _ P = 0 \eqno (3.12)
$$

Let's verify the selfconsistency condition (3.2). In our case for this
purpose it is enough to check up, that
$$
\langle \, z \, | \, (\, L _ nL _ {-n}
-L _ {-n} L _ n \,) \, | \, z \,\rangle _ P = 0 \eqno (3.13)
$$
The simple calculation shows:
$$  L_nL _ {-n} = \frac{1}{2} \, \sum^n_{m = 1} \,
m (n-m \,) + n \, (\alpha^1_0 \,)^2 + 2n \,\sum^{n}_{m = 1} \,
\alpha^1_{-m} \alpha^1_m +
$$
$$ + \sum^{\infty}_{m = n + 1}\,(n + m) \,
\alpha^1 _ {-m} \alpha^1 _ m + \sum^{\infty} _ {m = n + 1} \, (m-n) \,
\alpha^0 _ m\alpha^0 _ {-m} +
:L_{n} L_{-n}:                                      \eqno (3.14)
$$
Similarly:
$$
L_{-n} L_{n} = \frac{1}{2} \, \sum ^ {n-1} _ {m = 1} \, m (n-m \,) + n \, (\alpha ^ 0 _ 0 \,) ^ 2 + 2n \,\sum ^ {n} _ {m = 1} \, \alpha ^ 0 _ {m} \alpha ^ 0 _ {-m} +
$$
$$
+ \sum^{\infty}_{m = n + 1}\, (n + m) \, \alpha^0_{m} \alpha^0_{-m}
+ \sum^{\infty}_{m = n + 1} \, (m-n) \, \alpha^1 _ {-m}
\alpha^1_{m} + :L_{-n} L_{n}:       \eqno (3.15)
$$
Since \ $ :L_{n} L_{-n}: \equiv :L_{-n} L_{n}:$ \ from two last
equalities we obtain:
$$
L_nL _ {-n} - L_{-n} L_n  = 2n \, L_0 \,, \eqno (3.16)
$$
The ordering in the right hand side of Eq.(3.16) is given according to
(3.5). From (3.16) it is seen, that the equalities (3.12) take place, so
 here the selfconsistency conditions (3.2) are valid.

Note that, generally speaking,
$$
\langle \, z \, | \, L _ nL_{-n} \, | \, | \, z \,\rangle _ P\neq 0
$$
Let Hamiltonian of the system be \ $ {\cal H}_T = \sum _ n \, v_n \, L_n
$, where \ $ \bar {v}_n (\tau) = v_{-n} (\tau) $. Give the answer for the
question how the mean values of Heisenberg variables change in time. The
Heisenberg equations look like
$$
i \,\frac {d} {d\tau} \, \alpha^{\mu}_m =
m \,\sum _ n \, v_n \,\alpha ^ {\mu} _ {m + n} \,,
 \ \ \ \frac {d} {d\tau} \, x ^ {\mu}
 = \sum _ n \, v _ n \,\alpha ^ {\mu} _ n \eqno (3.17)
$$
Define by following equations the system of complex time-dependent
numbers:
$$
z^{\mu}_m (\tau) \, [\, k \,\delta (k '-k)\, ]
= \langle \, z ' \, | \,\alpha ^ {\mu} _ m (\tau) \,
| \, z \,\rangle _ P \,, \eqno (3.18)
$$
where \ $ \alpha ^ {\mu} _ m (\tau) $ \ are the Heisenberg variables.
In Eq.(3.18) we assume that the collections of parameters \ $ \{z \} $
\ and \ $ \{z^{\prime} \} $ \ are physical, that is they satisfy the
equations (3.10). Moreover \ $ z^{\prime\mu} _ m = z^{\mu}_m $ if \ $
z^{\prime\mu}_0 = z^{\mu}_0 $. Obviously in this case  \
$\langle\,z^{\prime}\,|\,z\,\rangle=k\,\delta(k'-k)$ . With the help of
(3.17) and (3.18) we obtain the equations for change in time
of the parameters (3.18):
$$
i \,\frac {d} {d\tau} \, z^{\mu} _ m (\tau) =
m \,\sum_n \, v_n (\tau) \, z^{\mu}_{m + n} (\tau) \eqno (3.19)
$$
Thus, the mean values of Heisenberg dynamic variables satisfy the
classical equations of motion.

Now consider the question about the Lorentz-invariancy of the theory.
Let's write the operator of the angular momentum in the form:
$$
J^{01} = (x^0p^1-x^1p^0 \,) +\frac {i} {2} \,
\sum _ {n\neq 0} \, \frac {1} {n} \, (\alpha^0_n\alpha^1_{-n} -
\alpha^1_n\alpha^0_{-n}\, )
   \eqno (3.20)
$$
From here we see that the action of the angular momentum operator on
physical states is well defined. Let
$$
| \, z ^ {\prime} \, \rangle _ P=(\,\exp-i\,\omega\,J^{01}\,)\,
|\,z\,\rangle_P
   \eqno(3.21)
$$
Since \ $ [L _ m, \, J ^ {01}\,] = 0 $ \ and operator \ $ J ^ {01} $ \ is
hermithian, the equations (3.12) and (3.13) take place as well for the
states (3.21). It means, that the states \ $ | \, z ^ {\prime} \, \rangle _
P $, determined according to (3.21) are physical states. Owing to (2.35)
the both states \ $ | \, z \,\rangle _ P $ \ and \ $ | \, z ^ {\prime} \,
\rangle _ P $ \ are the eigenstates for the momentum
operator \ $ p ^ {\mu}$. The scalar product
of physical states is conserved under the Lorentz
transformations (3.21):
$$
\langle\,z^{\prime}_1\,|\,z^{\prime}\,\rangle=
\langle\,z_1\,|\,z\,\rangle
$$

We shall calculate the mean values of dynamic variables variations
generated by the angular momentum operator. Let for any variable \ $ {\cal
O} $
$$
 \delta {\cal O}_{(L)} = -i \,\delta\omega \, [\, {\cal O}, \, J ^ {01} \,]
$$
Then for fundamental variables we have:
$$ \delta \, x ^ {\mu}_{(L)} =
\delta\omega \, (\eta ^ {\mu 1} x ^ 0-\eta ^ {\mu 0} x^1 \,) \,,
 \ \ \ \delta \,\alpha^{\mu}_{m(L)} = \delta\omega \, (\eta ^ {\mu 1}
\alpha ^ 0 _ m- \eta ^ {\mu 0} \alpha ^ 1 _ m \,) \eqno (3.22)
$$
For variations of mean values (3.18) we find:
$$
\delta \, z^{\mu}_{m(L)} = \delta\omega \, (\eta^{\mu 1} z^0_m-\eta^{\mu 0}
z ^ 1 _ m \,)         \eqno (3.23)
$$
We see, that under Lorentz
transformation the corresponding transformation of observable variables \ $
\{x^{\mu}, \, p^{\mu} \, \} $ \ occurs and also the gauge transformation of
 gauge degrees of freedom take place.

Let's discuss shortly the superposition principle under the second quantization method.

Assume, that the states \ $|\,z\,\rangle_P$ \ and $|\,z'\,\rangle_P$ are physical. Is the state

$$
|\,z\,\rangle_P+|\,z'\,\rangle_P
$$
physical?

We think, that {\it {it is not necessary to extend the superposition principle to unphysical, gauge degrees of freedom}}. Therefore, if
under the second quantization method the superposition principle
will appear limited in the space \ $H_G$ , in our opinion, it does not
depreciate the method. However, in physical space
the superposition principle is kept completely.

\centerline {}
\centerline {}
\centerline {\bf {4. Conclusion.}}
\centerline {}

 The quantization methods which was applied to the two dimensional string
in the Sections 2, 3 next we shall call as first and second
methods, accordingly. The results of this paper are
reduced to the following theses:

1) Both methods lead to absence of the central charge in Virasoro algebra.
In both cases the basis for such result is that in complete space of states
the scalar product is positively defined.

2) The first method successively realizes the Dirac quantization
prescription
for degenerated systems. In this case in the theory there are such
dynamic variables, which are not the operators in physical space of states.
In simple models (two-dimensional string and electrodynamics) this
fact is not important since the matrix elements of observables relative to
physical states are well defined and calculated obviously. However, in the
complicated theories such as four-dimensional gravity the specified
difficulty can extremely complicate the decision of a problem, by ceasing
to be in result only technical.

3) The second quantization method is based on some weakening of Dirac
quantization conditions for degenerated systems. It enables to treat the
all initial dynamic variables as the operators in physical space of states.
The physical space of states is a vector subspace of complete space of
states with positively defined scalar product. This statement is incorrect
for the first quantization method. The scalar product in physical space of
states is induced by scalar product in complete space of states.

4) The both quantizations result in isomorphic spaces of physical states.
However, it is unknown, whether this result will be kept in more
complicated models.

5) The both quantizations result in the Lorentz-invariant theories.
 Note that in the conventional quantization there exists a state which is
invariant under Lorentz transformations. This state is the ground state. In
this respect the conventional string theory is similar to the standard
quantum field theory of point objects. In such field theories the ground
state usually is Lorentz-invariant. Conversely, in our approach there does
not exist a state that is invariant under Lorentz transformations. For this
reason, the quantum string theory proposed above is analogous to a quantum
theory of a single relativistic particle. In the letter case
 there does not exist a
Lorentz-invariant quantum state of a single relativistic particle.
For existence of Lorentz-invariant state in our theory we would have to
introduce a second-quantized string field. In such
theory the ground state or vacuum would be Lorentz-invariant.

 It seems to us, that the second quantization method can give
interesting results in application to more complicated models.

\centerline { }
\centerline {\sc {Appendix}}
\centerline { }

 Here we describe the Dirac quantization [7] of free electromagnetic
Field. This quantization method is applied in Section 2.

The quantization of an electromagnetic field is presented in the form
$$
A_{\mu}(x) = \int \,\frac {d ^ 3k} {(2\pi) ^ 3} \, \frac {1} {\sqrt {2k ^ 0}} \,
\{a _ {\mu} ({\vec {k}}) \, e ^ {ikx} + a ^ + _ {\mu}
 ({\vec {k}}) \, e ^ {-ikx} \}
  \eqno (A1)
$$
Here $\mu ,\,\nu ,$ .... = 0, 1, 2, 3, \ $kx\equiv k_\mu x^\mu =-k^0x^0+
{\vec k}\cdot {\vec x},$ \ $k^0=\left|{\vec k}\right| $ and
 $\{a_\mu ({\vec k}),$ $a_\mu^{+}({\vec k})\}\,\,$
 are some generators of an associative involutive algebra
 ${\cal A}$ with an identity element (see Sec. 2). The nonzero commutation
relations between these generators have the form
$$
[ \, a _ {\mu} ({\vec {k}}), \, a ^ + _ {\mu} ({\vec {p}})\,] =
(2\pi) ^ 3 \,\eta _ {\mu\nu} \, \delta ^ {(3)} ({\vec {k}} - {\vec {p}})
  \eqno (A2)
 $$
One can see from the expansion (A1) that the set of elements $\partial _\mu
A^\mu (x)$ is linearly equivalent to the set of
elements $k^\mu a_\mu ({\vec k})$
and $k^\mu a_\mu ^{+}({\vec k})\,$ from the algebra ${\cal A}$. Let
 $a_i^T({\vec k})$ be two independent elements
  (for fixed ${\vec k}$) satisfying
the conditions
$$
\sum ^ 3 _ {i = 1} \, k _ i \, a ^ T _ i ({\vec {k}}) = 0 \,,
$$
$$
[ \, a ^ T _ i ({\vec {k}}), \, a ^ {T +} _ j ({\vec {p}})\,] =
(2\pi) ^ 3 \,\left (\delta _ {ij} -\frac {k _ ik _ j} {{\vec {k}} ^ 2} \right) \,
\delta ^ {(3)} ({\vec {k}} - {\vec {p}}) \eqno (A3)
$$
Eqs.(A1) and (A2) imply the following commutation relations
 \ $ (F _ {\mu\nu} = \partial _ {\mu} A _ {\nu} -\partial _ {\nu} A _ {\mu}) $:
 $$
[ \, F _ {\mu\nu}(x),\,k ^ {\lambda} \, a _ {\lambda} ({\vec {k}})\,] =
[ \, F _ {\mu\nu}(x),\, k ^ {\lambda} \, a ^ +_ {\lambda} ({\vec {k}})\,]=0 \,,
  \eqno (A4)
  $$
  $$
  [\, k ^ {\mu} \, a _ {\mu} ({\vec {k}}), \, p ^ {\nu} \, a ^ + _ {\nu}
  ({\vec {p}})\,] = 0
    \eqno (A5)
    $$
We have also
$$
[ \, a ^ T _ i, \, k ^ {\mu} \, a _ {\mu} ({\vec {k}})\,] =
[ \, a ^ T _ i, \, k ^ {\mu} \, a ^ + _ {\mu} ({\vec {k}})\,] = 0 \eqno (A6)
$$
Dirac quantization presupposes that the conditions
$$
a ^ T _ i ({\vec {k}}) \, \vert \, 0 \,\rangle = 0    \eqno (A7)
$$
are imposed on the ground state and the conditions
$$
k^ {\mu} \, a _ {\mu} ({\vec {k}}) \, \vert \ \rangle = 0 \,, \qquad
k^ {\mu} \, a ^ + _ {\mu} ({\vec {k}}) \, \vert \ \rangle = 0 \eqno (A8)
$$
are imposed on all physical states. As a result of Eqs. (A5) and (A6) the
conditions (A7) and (A8) are algebraically consistent. The states
satisfying the conditions (A8) are called physical. The Fock space of all
physical states is constructed with the help of the creation operators
 $a_i^{T+}({\vec k})$ from the ground state satisfying the conditions
  (A7) and (A8). As a result of Eq. (A6) any state of the Fock space
  constructed satisfies the conditions (A8).

Let $k_{-}^\mu =(-k^0,{\vec k})$. We find from Eq. (A2)
$$
[ \, k_{-}^{\mu} a _ {\mu} ({\vec {k}}), \, p ^ {\nu} \, a ^ + _ {\nu} ({\vec {p}})\,] =
2 \, {\vec {k}} ^ 2 \, (2\pi) ^ 2 \,\delta ^ {(3)} ({\vec {k}} - {\vec {p}})
  \eqno (A9)
$$
The relations (A4) and (A9) mean that the observables $F_{\mu \nu }$ do
not depend on the generators $\{k_{-}^\mu a_\mu ({\vec k}),$
 $k_{-}^\mu a_\mu ^{+}({\vec k})\}$.
Therefore all matrix elements of the form
$\left\langle \Lambda \right| F_{\mu \nu }\left| \Sigma \right\rangle $,
where $\left| \Lambda \right\rangle ,$ $\left| \Sigma \right\rangle $ are
 physical states, are determined.

We note that as a result of Eqs. (A3) and (A7) the scalar product in the
space $V$ is positive-defined provided that $\left\langle 0\mid
0\right\rangle =1$. We call attention to the fact that the action of the
generators $k_{-}^\mu a_\mu ({\vec k})$ and $k_{-}^\mu a_\mu ^{+}({\vec k})$
on the physical states is not determined in Dirac quantization, and
therefore these generators of the algebra ${\cal A}$ are not linear
operators in the space of physical states.

\centerline { }
\centerline {REFERENCES}
\centerline { }

\begin{itemize}
\item[1. ]
R. Jackiw, E-print archive, gr-qc/9612052.
\end{itemize}
\begin{itemize}
\item[2. ]
E. Benedict, R. Jackiw, H.-J. Lee, Phys.Rev. {\bf {D54}} (1996) 6213.
\end{itemize}
\begin{itemize}
\item[3. ]
D. Cangemi, R. Jackiw, B. Zwiebach, Ann. Phys. (N.Y.) {\bf {245}} (1996)
408;
D. Cangemi and R. Jackiw, Phys.Lett. {\bf {B337}},
271(1994); Phys.Rev. {\bf {D50}}, 3913(1994); D. Amati, S. Elitzur and E.
Rabinovici, Nucl.Phys. {\bf {B418}}, 45(1994); D. Louis-Martinez, J
Gegenberg and G. Kunstatter, Phys.Lett.  {\bf {B321}}, 193(1994); E.
Benedict, Phys.Lett.  {\bf {B340}}, 43(1994); T. Strobl, Phys.Rev. {\bf
{D50}}, 7346(1994).
\end{itemize}
%\vspace{-4mm}
\begin{itemize}
\item[4. ]
S.N. Vergeles. Zh.Eksp.Teor.Fiz., {\bf {113}}(1998) 1566.
\end{itemize}
\begin{itemize}
\item[5. ]
P.A.M. Dirac, Lectures on quantum mechanics.
Yeshiva University New York. 1964.
\end{itemize}
\begin{itemize}
\item[6. ]
V.B. Green, J.H. Schwarz, E. Witten. Superstring Theory. Cambridge
University Press, 1987
\end{itemize}
\begin{itemize}
\item[7. ]
P.A.M. Dirac. Lectures on quantum field theory. Yeshiva University,
 New York, 1967
\end{itemize}
\begin{itemize}
\item[8. ]
S.S. Schweber. An Introduction to Relativistic quantum field theory.
 Row, Peterson and Co. Evanston, N.Y., 1961.
\end{itemize}

\end{document}